\documentclass[acmsmall]{acmart}

\usepackage{soul}
\usepackage{pifont}
\usepackage{xspace}
\usepackage{subfigure}
\usepackage{tabularx}
\usepackage{svg}
\usepackage{varwidth}
\usepackage{booktabs}
\usepackage{tabularx}
\usepackage{enumitem}
\usepackage{longtable}
\usepackage{multirow}
\usepackage{array}

\newcommand{\SYSTEM}{PressProtect\xspace}

\newcommand{\myquote}[1]{\begin{quote}{\textit{#1}}\end{quote}}
\newcommand{\dimension}[1]{\textbf{\textit{#1}}}

\AtBeginDocument{%
  \providecommand\BibTeX{{%
    \normalfont B\kern-0.5em{\scshape i\kern-0.25em b}\kern-0.8em\TeX}}}

\setcopyright{acmlicensed}
\acmJournal{PACMHCI}
\acmYear{2024} \acmVolume{8} \acmNumber{CSCW2}
\acmArticle{509} \acmMonth{11} \acmPrice{15.00}
\acmDOI{10.1145/3687048}

\begin{document}

\title{PressProtect: Helping Journalists Navigate Social Media in the Face of Online Harassment}

\author{Catherine Han}
\email{catherinehan@cs.stanford.edu}
\affiliation{%
  \institution{Stanford University}
  \country{USA}
}

\author{Anne Li}
\email{anne24@stanford.edu}
\affiliation{%
  \institution{Stanford University}
  \country{USA}
}

\author{Deepak Kumar}
\email{kumarde@ucsd.edu}
\affiliation{%
  \institution{Stanford University}
  \country{USA}
}
\affiliation{%
  \institution{University of California San Diego}
  \country{USA}
}

\author{Zakir Durumeric}
\email{zakir@cs.stanford.edu}
\affiliation{%
  \institution{Stanford University}
  \country{USA}
}

\renewcommand{\shortauthors}{Han et~al.}

\begin{abstract}
  Social media has become a critical tool for journalists to disseminate their work, engage with their audience, and connect with sources. Unfortunately, journalists also regularly endure significant online harassment on social media platforms, ranging from personal attacks to doxxing to threats of physical harm. In this paper, we seek to understand how to make social media usable for journalists who face constant digital harassment. To begin, we conduct a set of need-finding interviews with Asian American and Pacific Islander journalists to understand where existing platform tools and newsroom resources fall short in adequately protecting journalists, especially those of marginalized identities. We map journalists' unmet needs to concrete design goals, which we use to build \SYSTEM, an interface that provides journalists greater agency when engaging with readers on Twitter/X. Through user testing with eight journalists, we evaluate \SYSTEM and find that participants felt it effectively protected them against harassment and could also generalize to serve other visible and vulnerable groups. We conclude with a discussion of our findings and recommendations for social platforms hoping to build defensive defaults for journalists facing online harassment.
\end{abstract}


%
%


\maketitle

\section{Introduction}
\label{sec:introduction}

Journalism has always been a dangerous profession. From 2016 to 2021, the United Nations Educational, Scientific and Cultural Organization (UNESCO) reported that 455 journalists were killed for their work or while on the job~\cite{unesco2021trends}. The online landscape of contemporary journalism has raised further concerns of novel threats to journalists' safety and expression, including doxxing, swatting, coordinated harassment campaigns, and dogpiling~\cite{unesco2022chilling,quodling2015doxxing}.
Such online violence disproportionately affects marginalized-identity journalists: in 2020, UNESCO found that 73\% of women journalists reported experiencing online violence in the course of their work~\cite{unesco2022chilling}. Likewise, the perceived lack of systemic support from news organizations and social platforms has burdened journalists, especially women and minority-identity journalists, with the responsibility of protecting themselves~\cite{holton2023problem,chen2020thickskin}. As a direct result of online violence, the constant vigilance required to monitor it, and the lack of resources to handle it, journalists have reported experiencing fatigue, burnout, or psychological trauma and have expressed a desire to exit the profession altogether~\cite{lee2023journalists,chen2020thickskin}.

Journalists are subject to various standards for social media engagement that have made existing online harassment resources difficult or impossible to use. Oftentimes both explicit and implicit, journalistic standards for social media require that journalists consume or even engage with potentially harmful content to do their jobs.
Not only do news organizations subject journalists to social media policies that encourage journalists to maintain an ``active'' and ``authentic'' online persona~\cite{nelson2021tightrope}, but also journalists' role in society obliges that they fulfill a social contract to inform the public~\cite{sjovaag2018journalism}.
Prior work has proposed systems to protect social media users more broadly against online harassment, leveraging community support to curate word or account blocklists~\cite{jhaver2018online,jhaver2022designing,blockpartyapp,block-together}. However, simply filtering content or accounts does not address journalists’ specific needs to monitor the comments they receive and remain accessible to the public. The tension between these professional standards for participation and unmet safety needs in online spaces has left journalists struggling to manage their online presence. 

In this paper, we begin by formalizing the specific online safety needs of journalists. We conduct semi-structured interviews with eight Asian American and Pacific Islander (AAPI) journalists to center the needs and concerns of a marginalized journalist group; our approach aligns with prior CSCW work that advocates for centering vulnerable populations to design technological systems that are more robust to abuse, better serving all users~\cite{blackwell2018online}. In our interviews, we focus on the nuances of their harassment experiences, protective strategies, and moderation practices. We coupled these interviews with a closed card sorting exercise to investigate journalists' decision-making processes when interfacing with harassment. We found that, at the time of our interviews in 2022, journalists heavily relied on social media platforms like Twitter/X for work, but these platforms were also the primary vectors for online harassment. Still, journalists expressed a high willingness to interact with users critiquing their stories and an appetite for understanding their audience response, even if their language was negative or passionate. As a result, journalists often manually weigh each of their comments, considering its relevance to their work and how reasonable the commenter appears, to decide how they might engage. 
Even when faced with abusive interactions, some journalists were hesitant to 
leverage existing platform moderation mechanisms (e.g., blocking or muting accounts). Because of the affordances of blocking on Twitter/X\footnote{Twitter/X indicates to a user when they have been blocked.} and the fundamental concept of blocking (denying access to view or interact with a user's content), journalists felt that blocking would not only empower their attackers as a signal of successful harassment, but also be antithetical to their duty to inform the public.

Building upon our findings, we design and evaluate \textit{\SYSTEM}, an interface for journalists to effectively use social media in the face of significant harassment.
\SYSTEM classifies online engagement along two axes: (1) the relevance of a comment to the journalist's post/story and (2) its toxicity. We use few-shot prompting of a large language model, GPT-3.5, to determine a comment's relevance to a given story. We combine this with Google Jigsaw's Perspective toxicity classifier~\cite{perspective} to determine if a comment is toxic.
The \SYSTEM interface displays non-toxic content by default, stowing ``relevant and toxic'' and ``irrelevant and toxic'' content behind additional user interface (UI) controls to give journalists agency in deciding when to engage with potentially harmful content.
For instance, while the ``relevant and toxic'' category may include offensive or inflammatory language, it may also contain information or feedback that is useful for journalists to process. Because \SYSTEM is a client-side interface that uses content-based features, it does not suffer from the drawbacks of platform-based account blocking that have made this mechanism unusable for some journalists.

To evaluate \SYSTEM, we conduct a user study with a separate set of eight journalists whose online harassment experiences on Twitter/X span a wide range, from receiving hate comments to being harassed by a stalker or targeted by a nation-state adversary. Through user testing, we explore how they interacted with content displayed through \SYSTEM's categorizations, feedback about its logic, and their reflections on incorporating such a system into their workflow. 
We find that \SYSTEM's logical abstraction of both the toxicity \textit{and} the relevance of comments provided a nuanced lens for viewing reader engagement that effectively protected journalists (especially marginalized-identity journalists or journalists reporting on controversial topics, such as climate change or immigration); journalists also felt that \SYSTEM~could generalize to serve other populations of users in the public eye (e.g., politicians, celebrities). Finally, through testing, we observe the emergence of a critical need for journalists: effectively flagging imminently dangerous threats to be addressed separately from other kinds of harassment.

We draw upon our observations from the design and evaluation of \SYSTEM to discuss their implications for social media platform design and opportunities for platforms, newsrooms, and third-party developers to collaborate on better protecting journalists. Our exploration of journalists' experiences and deficits in existing resources for online harassment illuminate various dimensions to their needs, distinguishing them from those of other at-risk populations. More broadly, we argue that both platforms and researchers consider the multi-dimensionality of needs across different vulnerable populations to design safer online experiences for diverse user populations.

\section{Related Work}
\label{sec:related_work}
Extensive literature spanning disciplines, including computer science, communication, and in the case of our work, journalism, has investigated various aspects of online harassment. We build on three primary bodies of related work to inform our design of \SYSTEM. First, we review the landscape of online harassment experiences for vulnerable communities (e.g., women, content creators, sex workers). Second, drawing upon journalism literature, we discuss the specific concerns of journalists facing online harassment. Finally, we review the different systems and tools that have been proposed to combat online harassment more broadly.

\subsection{Understanding community experiences of online harassment}
Online harassment is a common experience: as of 2021, 48\% of people globally have experienced harassment in some form~\cite{thomas2021hatesok}. Significant research has focused on the communities that experience disproportionately severe or frequent online harassment, including gender or religious minorities~\cite{nadim2021silencing,vitak2017identifying,powell2020digital,salehi2023sustained}, 
politicians~\cite{gorrell2018twits,hua2020characterizing,krook2020cost}, content creators~\cite{thomas2022s,samermit2023millions}, and sex workers~\cite{mcdonald2021s}.
Prior work has investigated the differences in harassment responses by gender, finding that targeted women are more likely than targeted men to manage their online presence by silencing themselves or blocking others~\cite{nadim2021silencing,vitak2017identifying}.\footnote{Similar to prior work on harassment and intimate partner violence, we intentionally avoid the term ``victim'' to not disempower people facing harassment~\cite{thomas2021hatesok, kumar2023understanding, bellini2023paying}.} Other research has suggested that politicians preemptively block highly adversarial users by analyzing their profile metadata~\cite{hua2020characterizing} and described how content creators felt that improved content removal mechanisms would effectively protect their personal and community safety~\cite{samermit2023millions}. Similarly, Salehi et~al.\ examine how visible Muslim Americans (e.g., journalists, activists, aspiring politicians) continue to engage online despite persistent harassment by leveraging protective platform affordances, such as blocking accounts or limiting who can engage with social media posts (e.g., only people you follow can reply to your tweet). However, they also discuss the fundamental limitations and negative externalities of content moderation for marginalized populations --- that is, that the harm is systemic and that content removal policies can be abused by attackers to further harass their targets.

Recent work from the CSCW community has taken an intersectional approach to studying online harms and moderation~\cite{gilbert2023towards}; prior work supports this direction by highlighting how intersectionality shapes users' harassment exposure and harm reduction tactics~\cite{wagner2022tolerating,krook2020cost}. Related, Schoenebeck et~al.~\cite{schoenebeck2023online} detail how harassment experiences of people in non-majoritarian (i.e., outside of North American and European) countries can help shape platform design that aligns with regional values. 
We draw upon this body of work to guide our understanding of journalists' online harassment experiences and inform our approaches to designing a system that caters to their unique online safety concerns. 

\subsection{Needs of journalists facing online harassment}
Journalistic reciprocity is an idea that situates journalists as community-builders that can directly, indirectly, and repeatedly over time, exchange with their readers to develop trust, connectedness, and social capital~\cite{lewis2014reciprocal}.
As social media has become the main medium for reciprocal journalism, online violence against journalists, particularly women, has emerged as a major threat to press freedom~\cite{lorenz-online-violence,lewis2020online}. Prior work has highlighted the threats journalists face by virtue of their occupation in an increasingly online world, including coordinated attempts to silence their freedom of expression, manufactured controversies to ruin career opportunities, and calls to action for their physical harm~\cite{holton2023not,tandoc2023digitization,ferrier2018trollbusters,chen2020thickskin,lorenz-online-violence}. Lewis et~al.\ explored what factors lead journalists to face more online harassment, such as gender and ``personal visibility,'' which is the degree to which personal traits (e.g., one's face or voice) are presented alongside or as a part of their work~\cite{lewis2020online}. 
However, journalists cannot simply retreat from the Internet. Online journalistic reciprocity has been normalized and often required as a part of their jobs~\cite{chen2020thickskin,nelson2023worse}. Because of this, the disproportion of harassment experiences has placed outsized burden on women journalists to decide if and how to respond to online harassment; while the most common protective action by journalists  facing online harassment is to stop engaging, this response further disadvantages marginalized journalists by silencing their voices or hindering their career development~\cite{lewis2020online}. Withdrawing from social media could also potentially alienate audiences; some journalists feel that responding to online attacks is essential to maintaining their journalistic integrity; and in some choice cases, journalists have even been fired for non-participation on social media~\cite{nelson2023worse}. As such, journalists have a unique relationship with social media where they must endure receiving online harassment because of their employers' policies. We explore these tensions and concerns through our need-finding interviews in Section~\ref{sec:interviews} and incorporate these findings into our interface design.

\subsection{Systems and tools to combat online harassment}
An expansive area of social computing literature has proposed many approaches for combating online harassment. Goyal et~al.\ introduced a framework to reason about these various protective systems across several axes: Prevention (precautionary measures), Monitoring (understanding harassment activities), Crisis (immediate response needs), and Recovery (mitigating impacts)~\cite{goyal2022you}.
Within Crisis and Recovery, systems like HeartMob~\cite{blackwell2018online} and Trollbusters~\cite{ferrier2018trollbusters} offer targets a sense of community and support during acute instances of online harassment. Other processes, such as the one proposed by Xiao et~al.~\cite{xiao2022sensemaking}, offer a restorative justice-inspired pathway to enable targets facing interpersonal online harm to make sense of their experiences. Goyal et~al.'s own tool, Harassment Manager, seeks to provide journalists, among other users, a streamlined method to document instances of online harassment~\cite{goyal2022you}.

Most relevant to our work are both academic research and deployed industry tools that focus on Prevention and Monitoring systems. 
Squadbox is one such tool that focuses on the application of ``friendsourcing'' --- the idea that harassing messages (in this context, e-mail) can be filtered and later reviewed by a trusted third-party to protect the user from firsthand exposure to harassment~\cite{mahar2018squadbox}. Recent work from Jhaver et~al.\ builds on these ideas in FilterBuddy, a system for content creators to create, share, monitor, and delegate the administration of word filters to combat harassment on their YouTube pages. Both Squadbox and FilterBuddy employ participatory design, a technique often used by CSCW and CHI researchers~\cite{ashktorab2016designing}. In addition to tools proposed by researchers, there have also been a variety of deployed industry solutions for online harassment. These tools range from community-curated account blocklists~\cite{block-together} to collaborative systems that use custom machine-learning classifiers~\cite{perspective} to aid human moderators in identifying abuse~\cite{jigsaw2016moderator}.

Tools like Block Party provide Twitter/X users with more fine-grained mechanisms for rule-based blocking of accounts, such as if an account has interacted with a problematic tweet or account or if it has a low follower count. These features can also be used to control notifications and the display of content from suspicious or harassing accounts~\cite{blockpartyapp}; however, the increasing tumult surrounding Twitter/X and the surge in enterprise API pricing have since forced out third-party tools for the platform, like Block Party~\cite{blockparty2023hiatus}. As we discuss in Section~\ref{sec:eval}, we identify components of these systems, such as friendsourcing, that some journalists employ. In Squadbox, participants were frustrated by the lack of agency in deciding if or when they would be exposed to harassing content. Likewise, in FilterBuddy, participants prioritized control so highly that they were willing to trade greater automation for more control over the tool's options. Further, the protective platform affordances (i.e., blocking and turning off comment access to non-friends on Facebook) that Salehi et~al.\ find are most preferred by users are also rooted in the desire to control who can engage with their accounts. Based on these findings, it is resoundingly clear that users want control over their online experiences --- both in how they encounter harmful content and in how they prefer to use protective tools. We explore the extent to which journalists are able to benefit from this suite of existing tools and use these observations in a participatory design process to develop a system that better addresses journalists' specific needs.

\section{Need-finding Interviews}
\label{sec:interviews}
To mindfully design an anti-harassment tool that meets the unique needs of journalists, we conducted semi-structured need-finding interviews with eight journalists (J1-J8) that self-identified as having experienced online harassment.
Through these interviews, we aimed to better understand journalists' experiences with and current strategies for handling online harassment. We also use the open-ended nature of these interviews to explore deficits in existing mitigation strategies or tools and how they might be addressed. We distill these observations into a set of needs (Section~\ref{sssec:needs}) that inform our interface design for a re-imagined and improved social media experience for journalists.

\subsection{Study Design}
\label{ssec:design}
We recruited participants from the Asian American Journalists Association (AAJA) convention in July 2022.
Following the suggestions of prior work in the HCI community~\cite{ashktorab2016designing,schoenebeck2023online,scheuerman2018safespaces,Blackwell2017classification,han2023hate}, we recruited from an organization that serves a minority identity in the United States (i.e., Asian American and Pacific Islander). By centering marginalized-identity journalists' experiences, we use a ``bottom-up'' approach as suggested by Blackwell et~al.\ to better design protections for all users~\cite{Blackwell2017classification}. 
We recruited using two primary methods: (1) we tweeted using the hashtag for the conference attendees, \#AAJA22, and (2) we posted in the conference Slack for all attendees. In both methods, we asked journalists to contact us (via e-mail or Twitter/X direct message) if they have had personal experiences with online harassment. We scheduled either in-person or video interviews with all of the journalists that reached out to us. 
Our recruitment efforts yielded eight participants, and while we acknowledge the limitations of our sample size (Section~\ref{ssec:limitations}), we concluded our need-finding recruitment when we felt we had reached adequate saturation~\cite{saunders2018saturation} --- that is, new themes did not emerge from interviews with additional participants --- to reason about journalists' harassment experiences and protective behaviors.
Although we did not collect other identity markers like sexuality or age, we collected details on their years of experience in journalism and the kinds of stories they cover.
Participants came from a wide range of professional backgrounds, from independent journalists to those in major newsrooms, and their coverage spanned various beats, or specialized reporting for a particular topic (e.g., politics, breaking news, sports); we report more detailed participant information in Table~\ref{tbl:interview_demographics}.
We formulated our interview questions to explore preliminary topics of interest, including how journalists use social media platforms professionally, what forms of online harassment they have experienced and how those experiences impacted them, and what protections they use for online safety and how they seek such resources. As participants spoke about their experiences, we prompted them to provide specific examples of incidents and asked follow-up questions to better understand the motivations of their actions or behaviors.
The full list of primary questions is included in Appendix~\ref{a:needfinding_qs}.

\begin{table}[h]
\small
\begin{tabular}{p{0.3\linewidth}p{0.3\linewidth}p{0.1\linewidth}}
\toprule
Demographic            &       Group                    &  N (out of 8)   \\
\midrule          
Beat                   &       Local News               &       2         \\
                       &       Climate                  &       1         \\
                       &       Investigative Reporting  &       1         \\
                       &       Breaking News            &       1         \\
                       &       Culture                  &       1         \\
                       &       International Relations  &       1         \\
                       &       College Football         &       1         \\
\midrule      
Years of Experience    &       1-3                      &       4         \\
                       &       4-6                      &       2         \\
                       &       7-9                      &       2         \\
\midrule      
Social Media Platforms &        Twitter/X               &       8         \\
Used for Reporting     &        Instagram               &       4         \\
                       &        Facebook                &       3         \\
                       &        WeChat                  &       1         \\
\midrule      
Gender                 &        Woman                   &       7         \\
                       &        Man                     &       1         \\
\bottomrule
\end{tabular}
\caption{\textbf{Demographics of need-finding participants}---%
    Demographic information of the eight participants who participated in our study.}
\label{tbl:interview_demographics}
\end{table}

\subsubsection{Mention Sorting Exercise}
\label{ssec:mentions_exercise}
To supplement our semi-structured interviews, we incorporated a closed card sorting exercise with each of the eight journalists we spoke to. The cards for this exercise were categories for how the journalist would prefer to interact with online reader engagement. We began this exercise only after the interview portion to avoid restricting participants' frame of mind during open-ended exploration. In this exercise, we curated a set of 10 abusive Twitter/X mentions that were previously sent to a prominent journalist on Twitter/X. We chose this set of mentions to span various forms of harassment, including personal attacks, critical article commentary, and sexually explicit content, to understand similarities or differences between how journalists process each of them. For every mention, we asked journalists to ``talk out loud'' as they sorted it and probed for more details when relevant to more concretely understand journalists' decision-making process for interacting with or moderating abusive engagement. The category cards were as follows:

\begin{enumerate}
    \item The tweet is displayed as normal; no moderation action is taken.
    \item The tweet is moved to a separate area before the journalist views it, as if in a spam folder.
    \item The tweet is removed from view by default; the journalist prefers to never have seen it.
\end{enumerate}

\noindent Interviews lasted between 20~minutes and an hour, with an average length of 35 minutes, depending on the degree of participants' exposure to harassment and their willingness to elaborate on their experiences. Participants were compensated with an Amazon gift code worth \$30.

After conducting the interviews, two researchers independently coded the transcripts into higher-level themes, such as ``social media platforms'' and ``mitigation strategy'' (Table~\ref{tbl:interview_codes}) using thematic analysis. The initial set of category labels were guided by the primary topics explored in the interview questions (Appendix~\ref{a:needfinding_qs}), but labels were iteratively added to the codebook when interview excerpts did not fit existing labels. Although we do not use these labels as a basis for describing our results quantitatively, we use inter-rater reliability (IRR) as a process to reduce confirmation bias, grounded in McDonald et al.'s guidelines for using IRR~\cite{mcdonald2019reliability}. Inter-rater agreement for this codebook was high, with a Kupper-Hafner agreement metric~\cite{kupperhafner1989irr} of 0.83. Both researchers met to resolve conflicts and obtain a final, coded interview dataset.

\begin{table}[h]
\small
\begin{tabular}{p{0.3\linewidth}p{0.6\linewidth}}
\toprule
Code                                &   Meaning \\
\midrule
Harassment trigger                  &   An event or action that triggers a flood of harassment from one or many users \\
Social media platforms              &   Discussion on what social media platforms they use and in what manner for their work \\
Tolerance threshold                 &   Factors behind the decision-making of whether or not to engage with users or moderate their content (e.g., severity, violent threat) \\
Harassment vector                   &   What vectors attackers use to harass journalists (e.g., e-mail, comments on social media platforms) \\
Kinds of hate                       &   Type of hate that was operationalized in an attack against them (e.g., racism, sexism, xenophobia) \\
Mitigation strategy                 &   Current strategies employed to combat harassment \\
Filtering \& moderating concern     &   Concern related to the effect or implication of filtering user content from their view or moderating users or their content \\
Resource-seeking strategy           &   Current strategies for seeking resources or community support for online harassment \\
\bottomrule
\end{tabular}
\caption{\textbf{Codebook for need-finding interviews} ---%
    Codebook for categorizing the different themes that emerged from our need-finding interviews with eight journalists.}
\label{tbl:interview_codes}
\end{table}

\subsubsection{Ethical Considerations}
\label{sssec:ethics}
We interviewed journalists from minority communities about potentially triggering experiences with online hate and harassment. Because of the sensitive nature of this material, we fully informed participants of the purpose of the research and the kinds of questions we planned to ask before beginning the interviews. We received verbal consent before beginning the interview and reminded participants that they could decline to answer any questions or stop the interview at their discretion. We removed any potentially personally identifiable information (PII) from any quotes we included in this paper. This research was approved by our Institutional Review Board (IRB).

\subsection{Interview Findings}
\label{ssec:needfinding_results}

In this section, we present the results from our exploration of how journalists experience online harassment, what journalistic norms govern their online interactions, and what strategies journalists employ to protect themselves. We synthesize these findings into a set of currently unmet safety needs and use this to guide the design of a protective interface that helps restore agency to journalists in the face of significant harassment (Section~\ref{sec:system}).


\subsubsection{Characterizing journalists' online harassment experiences}
\label{sssec:experiences}
\paragraph{Various threat models for online harassment}
We begin by investigating how journalists experience online harassment. Six participants described attacks on their journalistic credibility or integrity as a major form of harassment. Consistent with intersectionality theory~\cite{crenshaw1991mapping,collins2016intersectionality}, all of our participants' harassment experiences as journalists were shaped by various facets of their identity, including their race, gender, sexuality, and nationality. For instance, J5 recounted an experience where they were accused of conspiring with a foreign government agenda because of their foreign nationality as a journalist reporting on international politics in the United States. 
We next sought to understand what might trigger harassment. Seven of the eight journalists pointed to the release of a new story as the main entry point for online attacks, but the attacks they experienced varied significantly in burstiness and duration. Several journalists described attack patterns where they would receive a sudden flood of emails or replies on Twitter/X in response to the release of a story. Others dealt with persistent attackers that continued to harass them long after the initial story of interest, tracking the journalists' various accounts and emails even as they moved to different newsrooms during the course of their career.
When asked which vectors were most commonly used for harassment, over half of our participants mentioned Twitter/X; other platforms included more direct, private forms of communication, including WhatsApp messages, Facebook messages, text messages, and e-mails. Four journalists pointed to replies to their tweets as the most common vector for harassment. 
Two journalists experienced more coordinated, cross-platform harassment campaigns (J3, J7). J3 recounted that a ``handful of ringleaders'' directed users on one platform (Reddit) to execute attacks across their accounts on other platforms:

\myquote{
    ``I experience harassment on Twitter, Instagram, and Reddit, but I think I miss a lot of what happens on Reddit because I choose not to look at it. A lot of it happens on Asian American, misogynist subreddits, where\ldots they talk about me by name with some frequency. Sometimes, I think people that are either active on Reddit or Twitter would find [my personal Instagram] and tag me in hateful posts or comments that are quite abusive.'' -- J3
}

\noindent
The characteristics of these more complex attacks, such as their motivation, coordination, and execution across various platforms, parallel the experiences of other at-risk communities. For instance, Jhaver et al. examined how brigading, a coordinated attack by one online group on another, often occurs on Reddit and Twitter/X due to opposing ideologies and regularly targets marginalized groups (e.g., women, abuse victims)~\cite{jhaver2018online}. Prior CSCW research has also investigated the properties of hate raids, which are attacks that flood a target's chatroom with hateful messages, on the livestreaming platform Twitch~\cite{han2023hate,cai2023hate}. In more severe cases, these hate raids can also be coordinated on or spread to other platforms. Furthermore, both J3 and hate-raided streamers felt that their marginalized identities (e.g., gender, race, sexuality) motivated their attackers to select them as targets. Therefore, mirroring the characteristics of previously studied attacks on other platforms, the threat model described by J3 and J7 situates itself within what Marwick describes as morally motivated networked harassment: that is, harassment that leverages the amplification of a cross-platform, networked audience ``justified'' by the target's violation of identity norms~\cite{marwick2021networked}.

\paragraph{Extending beyond psychological harms}
In addition to experiencing cross-platform, coordinated attacks, half of our participants detailed experiences where online harassment escalated to offline harms, which have been been well-documented in the literature~\cite{lewis2020online,holton2023not,tandoc2023digitization,ferrier2018trollbusters,chen2020thickskin,lorenz-online-violence}. These harms threatened participants' physical safety through doxxing and violent language, which two participants described as sexually violent in nature. One journalist discussed how a well-resourced and motivated adversary, such as a nation-state, can deeply complicate these threats, even endangering the physical or financial safety of their loved ones.
%
%
Another journalist (J7) detailed how they became a target of harassment by someone who was initially a source for a story. J7 explained that they were accused of racism because of a miscommunication with a representative of an organization they were interviewing, which led to a coordinated attempt by that organization to harm J7's reputation. These public accusations were conducted across several of J7's social media accounts --- both personal and professional --- and cost J7 a job opportunity.


Informed by prior work, we find that our participants' concerns over these threats are shared by other vulnerable populations online, such as content creators~\cite{thomas2022s} and visible Muslim Americans~\cite{salehi2023sustained}. All of these populationsacknowledge that while threats to their physical and financial safety occur less frequently than other forms of harassment, the impact of such threats makes them an utmost concern, especially for women within these groups.

\subsubsection{Understanding journalistic norms centering online engagement}
\label{sssec:norms}
Despite online harassment and its resulting harms, journalists cannot withdraw altogether from social media. 
Aside from the importance of press freedom, our interviews resoundingly confirmed that social media platforms are an integral part of journalists' professional workflows. In fact, the platforms that journalists listed as primary vectors of harassment were also the platforms they used most for work; all of the journalists we interviewed stated that Twitter/X was the primary and most important social media platform for their profession.
While prior work has found that public-facing platforms like Twitter/X are often more hostile ecosystems for journalists than more private platforms like Facebook~\cite{lewis2020online}, we find that the public nature of Twitter/X makes it most central to journalists.
Our participants detailed various professional uses for social media, including keeping up to date with breaking news, reaching out to sources, and promoting their work. J6 described this ``need to have a social media presence'' as a double-edged sword that opened a communication channel between journalists, sources, and readers, but ultimately left journalists vulnerable. 

Many of our participants felt they needed to actively engage in online discourse. According to their responses in the mention sorting exercise (Section~\ref{ssec:mentions_exercise}), two valuable aspects of engaging with their online audience were (1) understanding what opinions people have and why, and (2) refuting unverified or incorrect information in the comments. For the latter, two of our participants explained that they felt responsible as journalists for curating comments on their social media posts so that readers consume accurate information. 
Our participants also acknowledged that journalistic norms require that they appear receptive to criticism. Three journalists noted that they hesitate to block attackers or hide problematic replies, as both of these are publicly visible actions on the Twitter/X platform. As such, journalists wishing to use these features face potential backlash: accusations of censoring free speech or demonstrating poor journalistic professionalism. Several journalists elaborated on how silencing themselves or the dissenting opinions of others on social media was fundamentally at odds with their perceived role as public informants:

\myquote{
    ``If people want to criticize me, it still is in the public interest for people to be able to see my work. People should be able to follow me and look at my work and look at my tweets, because I'm a journalist, and I serve the public.'' -- J2
}

\myquote{
    ``[As a journalist] you do want to be receptive to feedback, because you are doing this to inform an audience. You are not in an echo chamber. I am not a blogger.'' -- J4
} 

We contextualize these observations through the concept of \textit{reciprocal journalism}. Lewis et~al.\ argue that reciprocity, or the principal of mutual exchange within a community, applied to journalism serves as a mechanism for journalists to re-imagine a journalist-audience relationship that positions journalists as community-builders. While Lewis et al. acknowledge the possible harms of social media exchanges, such as revenge or trolling~\cite{lewis2014reciprocal}, other research has found that journalists experience cognitive dissonance from the harassment they receive when engaging in reciprocal journalism; despite this cognitive dissonance, researchers found that organizational and personal influences that encourage reciprocity were too strong for any of the journalists they studied to stop reciprocating altogether~\cite{deavours2023reciprocal}. Consistent with these results, we find that journalists are subject to professional expectations that make the passive consumption of, active engagement with, and curation of social media content central to their work, ultimately requiring journalists' participation in online spaces. 

\subsubsection{Mitigation strategies}
\label{sssec:mitigation}
\paragraph{Limited use of third-party \& platform moderation tools}
Although all of our participants described needing online protections, they largely did not rely on third-party tools to defend against online harassment. Only two journalists mentioned such tools, naming Block Party (an anti-harassment tool for Twitter/X that facilitates bulk blocking accounts)~\cite{blockparty2023hiatus} and Circleboom (a Twitter/X management tool that provides analytics and tweet deletion services).
We found that journalists' limited usage of third-party safety tools is due to the ``lack of awareness of any new resources and third party tools'' (J7) and the functional limitations of existing tools, which we detail later in this section.
Although J7 did not use third-party tools themself, they felt that third-party tools had the potential to democratize protections against online harassment for local newsrooms and independent journalists, who do not have access to the same resources as larger newsrooms.
One participant described bespoke security and safety filters for blocking phone calls offered by their large newsroom, but none of our participants mentioned their organizations when describing how they learn about new resources or seek support in the face of harassment.

As far as first-party safety and moderation mechanisms offered by social media platforms, five journalists mentioned using the blocking or muting features on Twitter/X. However, three of these five noted that blocking is not a mechanism that they often leverage. One journalist explained that they would only consider escalating to blocking or muting if the attacker was persistent. 
In the mention sorting exercise, we found that journalists' preferences for removing content greatly varied. However, the type of inflammatory comment that seven of the eight journalists opted to remove was sexually explicit language directed toward a journalist; two of these seven journalists explained that this was because the sexual comment bordered on physical threat. We also observed that if the comment inquired for the journalist's PII, threatened their physical safety, demonstrated a pattern of ``bad behavior'' from an account, contained sexually explicit language, or included unverified or incorrect information, journalists were more likely to take actions such as blocking, reporting, and muting.

\paragraph{Pitfalls of Account Blocking}
\label{para:blocking_pitfalls}
We next investigate our participants' general apprehension of blocking, one of the main protective mechanisms provided by social media platforms. We find that some journalists believe that blocking offending accounts is too costly in terms of both time and mental well-being. J2 elaborates that because of these costs and how they typically receive harassment, blocking was not a viable option for them:

\myquote{``[Blocking is] just not worth the effort. A lot of the times, I feel like the harassment is more than likely coming from a different person, or at least different accounts, each time. So, I don't think blocking someone will do much.'' -- J2}

\noindent Block Party attempts to reduce the cost of blocking by allowing its users to block others in bulk based on account attributes, such as number of followers, account age, profile description, or engagement history with tweets of interest (e.g., blocking all users that retweeted or liked a tweet). However, even bulk blocking raised concerns for our participants: two participants worried that blocking accounts with few followers could obstruct constructive dialogue or even access to valuable sources. J3 described the frustrations that they had when using Block Party:

\myquote{``I have some settings in [Block Party] about the types of account I don't want to hear from, but I think they do end up catching `friendlies' that are simply smaller accounts. I find that frustrating because I was a small account, and I want to hear from people that I am not fighting with.'' -- J3}

\noindent As such, J3 felt that existing third-party tools did not cater to journalists' specific needs, stating that if they were to use a tool, its design must show that ``it understands the actual experiences that people who get harassed go through and address their concerns and fears,'' rather than forcing them to ``shoehorn'' their situation into its protective mechanisms. 

Furthermore, while blocking can shut down persistent behavior from a specified account, it still fails to protect against the use of ``sockpuppet'' accounts --- accounts that are a part of a set of many accounts controlled by a single ``puppetmaster''~\cite{kumar2017sockpuppet}, or a distributed attack model spanning numerous legitimate accounts. Beyond the technical efficacy of blocking, our participants were also wary of its negative externalities. For instance, two journalists described a threat model in which they received harassment from a colleague; however, they felt that the complexities of their social and professional networks, paired with the public visibility of blocking and muting on Twitter/X, made using these features too detrimental to their professional development. Additionally, J8 described how blocking could give attackers validation that they harmed their targets: 

\myquote{
    ``It feels like you are giving something up when you choose to block somebody, and they can see that they got under your skin. I would worry that it would cause other people to [harass you] too.'' -- J8
}

\noindent
As such, J8 felt that blocking their attackers could actually encourage them and trigger further harassment, creating a positive feedback loop. While not described by our participants, prior work~\cite{salehi2023sustained} also discusses the weaponization of protective platform affordances, such as the mass reporting of target accounts or content as a silencing tactic by attackers, which has restricted vulnerable populations usage of such tools.

\paragraph{Raising thresholds for social media exchanges}
\label{para:raising_thresholds}
Due to the shortcomings of existing protection mechanisms in the face of reciprocal journalism norms, half of our participants stated that they typically try to ignore online harassment rather than actively moderate it.
Across the board, we found that journalists have raised their thresholds for actively engaging with their audience using various criteria, such as the account attributes of a commenter or the perceived intention or usefulness of a comment, to inform their decisions. For example, one factor that would encourage journalists to actively exchange with a reader was if their account appeared legitimate (e.g., they had a profile picture and username that seemed to correspond with their real identity, they had a reasonable number of followers, they had a history of platform usage). Likewise, three journalists described that they were more willing to interact with their audience if the exchange was a critique of their work. On the other hand, we found that journalists also justified ignoring or dismissing comments if they felt they were ``not to be taken seriously'' or the language undermined the commenter's own credibility (e.g., the commenter is perceived as irrational or unintellectual when they use baseless ad hominems). These observations echo prior work that has documented journalists' increasingly negative perception of online audience interactions~\cite{lewis2020online}. In the closed card sorting exercise, journalists generally did not opt to remove such irrelevant or nonsensical comments, instead preferring to passively view (and dismiss) them. Particularly, two journalists specified that they would still want to see even offensive or rude comments that engaged with the topics in their story, even if they were not insightful or polite. Six of the eight journalists opted to defer reading, or separate as if in a spam folder, at least one of the inflammatory comments. These journalists preferred deferring the view of inflammatory comments to removing them when they felt that viewing such a comment would allow them to better understand their audience or monitor patterns of harassment from repeat attackers for escalation potential.

\subsubsection{Summarizing needs}
\label{sssec:needs}
Above, we discussed the various threat models of online harassment that journalists face, the professional norms that dictate how journalists operate in online spaces, and the mitigation strategies that journalists employ (Section~\ref{sssec:mitigation}). Based upon these findings, we argue that the tension between the expectations of reciprocal journalism and journalists' desire to protect themselves from harassment raise online safety needs that are unaddressed by current mitigation strategies. 
Grounded in our interviews, we summarize the following outstanding needs of journalists when engaging with online audiences:

\begin{itemize}
    \item Journalists need to use social media to contact sources, receive tips, and gauge reader feedback.
    \item Journalists need to actively participate on social media platforms to engage in reciprocal journalism and adhere to professional norms.
    \item Journalists need protection from several different types of online attacks, including one-time commenters, persistent abusers, and large-scale networked harassment.
    \item Journalists need protections that do not validate the efforts of attackers or trigger further harassment.
\end{itemize}

These needs reflect that while journalists must actively use social media, they are not equipped with the tools or resources to do so safely. 
As we described in Section~\ref{sssec:mitigation}, news organizations do not adequately provide resources or support to journalists facing online harassment. This gap has left journalists to turn to third-party tools or platform-based levers to protect themselves. However, these protections ultimately fail to resolve the tension manufactured by journalists' need to cultivate bidirectional journalist-audience relationships. We argue that these deficits make existing protection mechanisms difficult, undesirable, or impossible for journalists to use effectively.

\section{\SYSTEM: Navigating Social Media in the Face of Harassment}
\label{sec:system}

Derived from the unmet needs that we identified in Section~\ref{sssec:needs}, we synthesize how we can make social media safer and more usable for journalists in the face of online harassment. In this section, we begin by defining design goals for such an interface. We then use these goals to guide our design of \SYSTEM, which provides journalists greater control over engaging with readers on Twitter/X by adding UI controls for the display of different types of reader replies. 

\subsection{Design Goals}
\label{ssec:designgoals}

We map the needs in Section \ref{sssec:needs} to the following concrete design goals (G1--G6):
\begin{enumerate}[label=(G\arabic*)]

    \item Journalists should be able to seamlessly access helpful and harmless reader responses while protecting themselves from harmful reader responses.
    


    \item Journalists should still be able to interact with harmful responses that are useful for their work, without simultaneously encountering harmful and unhelpful responses.

    \item Journalists should be able to access and understand the full scope of reader responses to their articles, including those that are harmful, if they so choose.

    \item Journalists should be able to use these protections even in the face of large-scale networked harassment like dogpiling\footnote{On social media, dogpiling is when someone is targeted by large groups ~\cite{marsden2017pileon}.} or adversaries that leverage bots or sockpuppets.

    \item Journalists should be able to use protections without the fear of silencing good-faith users.
    
    \item Journalists should be able to use protections without the fear of affirming attackers' attempts to harass them or appearing to censor free speech or ignore feedback.
            
\end{enumerate}


\subsection{From Goals to Interface}
\label{ssec:interfacedesign}

Guided by these design goals, we introduce \SYSTEM, an interface for Twitter/X intended to make the platform more usable for journalists. We chose to develop our interface for Twitter/X, because based on our need-finding interviews, it is the most central platform to journalists' work. Furthermore, we focused on replies to journalists' stories that are posted on Twitter/X because journalists identified replies as a primary vector for harassment. Motivated by journalists' need to use social media to gauge reader feedback and engage in reciprocal journalism (Section~\ref{ssec:needfinding_results}), \SYSTEM allows journalists to decide when to consume or interact with different types of reader replies. 
Based on G1--5, it categorizes comments using content-based attributes (i.e., how harmful and helpful a comment might be) to proxy journalists' process for reasoning about reader engagement.
Using these categorizations, \SYSTEM adds UI controls for content that could be harmful, guided by G1 and G2. 

This content-based abstraction for reasoning about reader replies also naturally handles threat models that involve many accounts, such as dogpiles and sockpuppets, because it does not rely on account attributes (G4). Similarly, this content-based approach satisfies G5 because it does not deny smaller or new accounts access to engaging. However, grounded in G3, the interface does not dispose of any content, even if it is harmful: journalists can still choose to view all comments. Finally, the protections offered by the interface are fully client-sided, as required by G6. Commenters, even abusive ones, are not aware of the journalist's use of \SYSTEM or how their comments are categorized, and they are not denied access to engaging with the journalist or their work. Below, we discuss the implementation of \SYSTEM's two underlying components: how content is categorized and how these categorizations are presented to the journalist. 

\subsection{Content Categorization} 
Accomplishing G1 and G2 hinges on emulating journalists' processes for reasoning about how harmful and helpful reader responses are.
To do this, we create working definitions of \textbf{toxic} and \textbf{relevant} content, grounded in the results of our need-finding interviews (Section~\ref{para:raising_thresholds}). We define \textbf{toxic} in accordance with Google Jigsaw's Perspective, an API that classifies comments by toxicity using machine learning, which states the following: ``We define toxicity as a rude, disrespectful, or unreasonable comment that is likely to make someone leave a discussion."\footnote{https://perspectiveapi.com/} This definition of toxicity maps directly to content attributes that shape journalists' willingness to engage with reader comments, such as if the comment appears irrational. Because journalists derive utility from comments that engage with content or topics in their stories and are more likely to engage with such comments, \SYSTEM uses how relevant a comment is to their stories as a proxy for how useful it is. Therefore, we define a \textbf{relevant} comment as one one that addresses the content of the news article linked in the journalist's tweet, as determined through a GPT-3.5 prompting process that provides both the article and comment text.

\begin{figure}[h]
    \hspace{-5em}
    \includegraphics[width=0.48\linewidth]{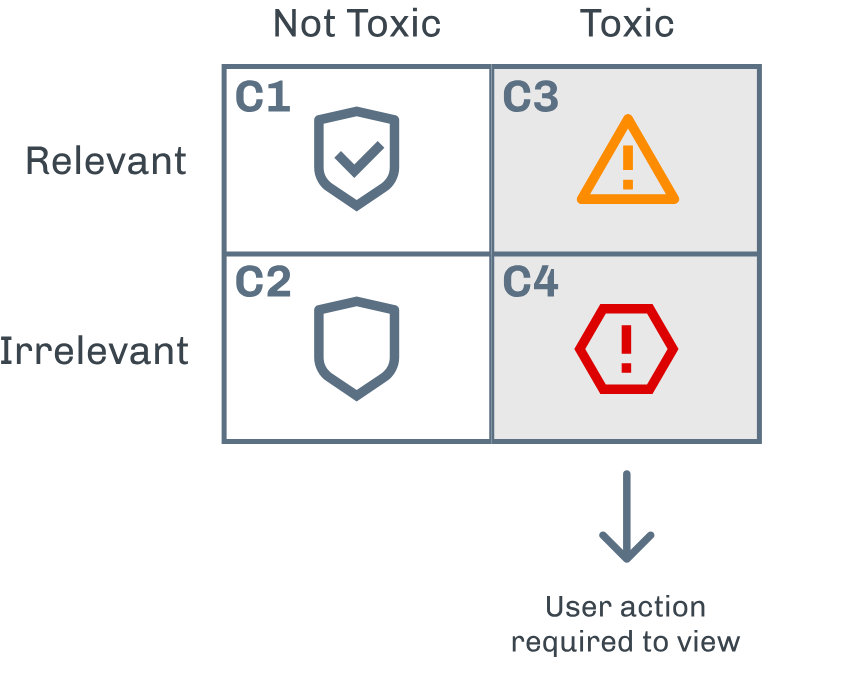}
    \caption{\SYSTEM classifies replies according to their toxicity and relevance to the journalist's article, as we determined that these axes reflect how journalists often reason about engagement. Journalists can seamlessly access C1 (relevant and not toxic) replies, and C2 (irrelevant and not toxic) replies are also quickly accessible. \SYSTEM provides additional UI protections when exposing C3 (relevant and toxic) and C4 (irrelevant and toxic) replies.}
    \label{fig:category_quadrants}
\end{figure}

Using our toxicity and relevance classifications, we bucket comments into four categories (C1--C4), as shown in Figure~\ref{fig:category_quadrants}:

\begin{itemize}
    \item \textbf{C1}: \emph{Relevant and non-toxic} --- replies that may be useful to the journalist with minimal harm. These might include productive discussions on the quality of the reporting or reference future directions, tips, and sources.
    \item \textbf{C2}: \emph{Irrelevant and non-toxic} --- replies that may be less useful to the journalist, but are likely harmless. For example, they could be lighthearted or tangential remarks.
    \item \textbf{C3}: \emph{Relevant and toxic} --- replies that are possibly harmful, but could contain information that is useful to the journalist. For example, they might present commentary on something the journalist missed but be couched in offensive language.
    \item \textbf{C4}: \emph{Irrelevant and toxic} --- replies that are likely harmful and not useful to the journalist. For example, they may be hate comments with little substance.
\end{itemize}

\noindent We provide examples of comments that a journalist might encounter in each of these categories in Table~\ref{table:example_comments}. We paraphrase these example comments from real comments in participants' data to preserve anonymity.

\begin{table}
\centering
    \begin{tabular}[h]{ | *1{l|} m{0.4\linewidth} | m{0.4\linewidth} | }
      \hline
      \ & Not Toxic & Toxic  \\ 
      \hline
      \multirow{1}{*}{\rotatebox[origin=c]{90}{Relevant}}  & ``I wouldn't have ranked X over Y, but I can see why you did.'' & ``They did terrible things, yet you wrote a glowing story. What a disgrace.''  \\ 
      \ & ``Thanks for your article, but you used the wrong term for this illness.'' & ``This is such a stupid ruling.''  \\ 
      \hline 
      \multirow{1}{*}{\rotatebox[origin=r]{90}{Irrelev.}} & ``Geez'' & ``No one gives a ****, this is 2023.''  \\
      \ & ``@USERNAME'' & ``Astrology is just as fake as climate change.''  \\ 
      \hline
    \end{tabular}
    \vspace{3pt}
    \caption{\textbf{Example Comments}---%
        Comments that might fall into each content category (C1-C4).}
    \label{table:example_comments}
\end{table}

\subsubsection{Toxicity}
To tag tweet replies that may contain harassment, we use Google Jigsaw's Perspective API~\cite{perspective}, which is a state-of-the-art toxicity detection system that has been used extensively in prior work~\cite{kumar2023understanding,hua2020characterizing,saveski2021structure,xia2020exploring}.
For the purposes of our tool, we use Perspective's \texttt{TOXICITY} classifier with a threshold of $0.5$ to identify if a comment is toxic. We use this threshold, as prior work observed this threshold~balanced precision and recall well for Twitter/X content~\cite{saveski2021structure}. We confirm this with our own thresholding experiment using a dataset from Kumar et.~al~\cite{kumar2021designing}, which we detail in Appendix~\ref{sec:threshold}.

\subsubsection{{Relevance}}
To determine whether a tweet reply is relevant to a given article, we use few-shot prompting of OpenAI's GPT-3.5, an off-the-shelf large language model (LLM). To evaluate whether this approach fit our task well, we first built a ground truth corpus of relevant tweet replies for a set of published articles. To do this, we collected 1,787~tweets by 1,439~journalists from eight different newsrooms: The Wall Street Journal, AP News, Financial Times, The Guardian, Bloomberg News, The New York Times, The Washington Post, and USA Today. We only considered tweets that contained a link to a published news story from each outlet. We scraped the content of each news article and processed each one using the Newspaper3k library,\footnote{\url{https://newspaper.readthedocs.io/en/latest/}} which provided us both the title of the article and a sanitized version of the article text. We then collected each tweet reply to the original tweet containing the link. In total, we amassed 3,973 tweet~replies. Two independent researchers then coded a sample of 300~tweet replies after reading the article text to establish a ground truth test set for tweet reply relevance to news articles. There was high agreement between the raters, achieving a Cohen's Kappa score of 0.8. The raters then resolved differences and ultimately aligned on a final ``ground truth'' corpus.

We attempted three strategies to identify tweet relevance: keyword matching based on article title, topic-matching using Latent Dirichlet Allocation (LDA), and few-shot prompting of GPT-3.5. For the LLM evaluation, we use the following prompt:

\begin{center}
\begin{varwidth}{\linewidth}    
\begin{verbatim}
The following is the text of a news article:
<article_text>.

Consider the following comment:
<comment_text>

Return a JSON object with a field, "relevance," that is a 
score from 1 to 5 depending on how relevant the comment 
is to the article.

\end{verbatim}
\end{varwidth}
\end{center}

\noindent We then provide three examples of relevant and irrelevant replies to better scope the LLM to our task. We consider a score of three or higher to be ``relevant'' and a score below three to be ``irrelevant.'' We note that while there may be more effective prompting strategies to improve the performance of LLM~reasoning for this task (e.g., Chain-of-Thought prompting or self-reflection), the purpose of our work is to understand how applicable \SYSTEM's abstraction of toxicity and relevance is to journalists' usage of social media. 

\begin{table}
    \centering
    \small
    \begin{tabular}{lrrrr}
        \toprule
        Relevance Technique       &   Precision &   Recall   &   Accuracy    &   F1 \\
        \midrule
        Title Keywords  &   0.91        &   0.57        &   0.59        &   0.70 \\   
        LDA             &   0.95        &   0.12        &   0.25        &   0.21 \\
        GPT-3.5         &   0.86        &   0.75        &   0.74        &   \textbf{0.80} \\
        \bottomrule
    \end{tabular}
    \vspace{3pt}
    \caption{\textbf{Techniques to Identify Relevant Tweets to Article Text}---%
        We evaluate three different techniques to identify if a candidate tweet reply is relevant to the text of an article: a keyword-based approach, latent Dirichlet allocation (LDA), and an off-the-shelf LLM (GPT-3.5). LDA performs the worst at this task, achieving an F1 of just 0.21, compared to 0.7 from title keyword matching and 0.8 from GPT-3.5.
    }
    \label{table:relevance_scores}
\end{table}

Table~\ref{table:relevance_scores} shows how well each strategy performs on our test set in terms of precision, accuracy, and F1. Although LDA provides the highest precision, we find that it suffers from significantly low recall and heavily skews towards labeling replies as ``irrelevant.'' In contrast, both title keyword match and GPT-3.5 achieve high F1~scores, with GPT-3.5 achieving an acceptable F1 of 0.8, suggesting a good balance between precision and recall on this task. As such, for the scope of our tool, we utilize GPT-3.5 for identifying tweets that are relevant to article texts in our system.

\begin{figure}
    \centering
    \includegraphics[width=\linewidth]{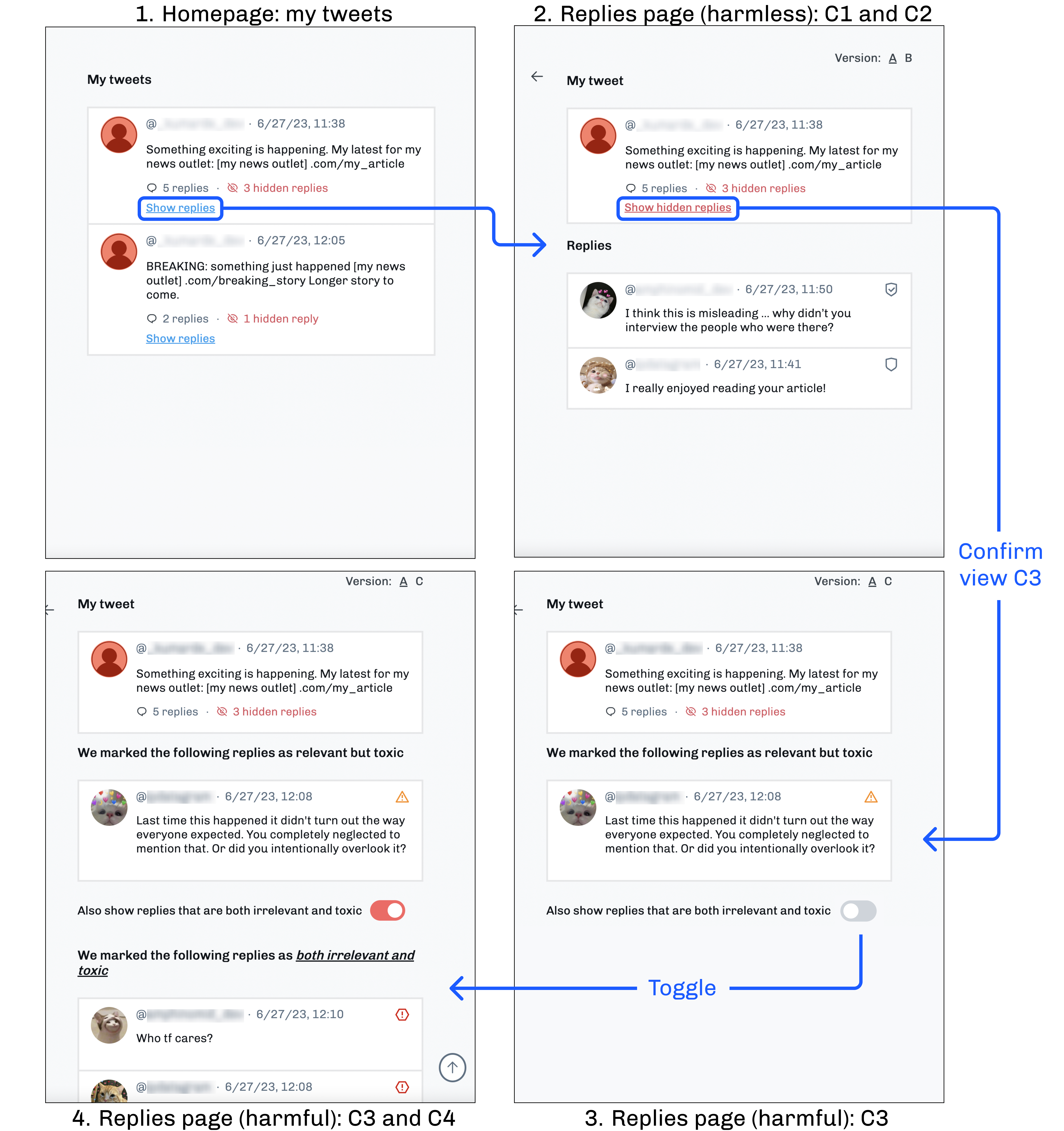}
    \caption{\SYSTEM presents comments differently based on their categorizations, providing additional UI protections to the journalist for content that could be harmful. Panel 1 shows the \SYSTEM homepage, which displays the journalist's tweets. Panel 2 shows the page for harmless replies, grouping relevant replies at the top (C1 and C2 visible). Panel 3 shows the page for harmful replies, with a toggle that protects the journalist from being exposed to harmful and irrelevant replies (C3 visible). Panel 4 mirrors Panel 3, but with the the toggle enabled and the irrelevant replies displayed under the relevant ones (C3 and C4 visible). These UI protections enable the journalist to navigate the different categories in a safe and controlled way.}
    \label{fig:exploratory_design}
\end{figure}

\subsection{Content Presentation}
To provide journalists safety and support in their engagement with potentially harmful content, we display comments to journalists differently based on their content categorizations. Our presentation of the comments is guided by G1-3, which states that journalists should be able to seamlessly access helpful, harmless reader responses they receive while being protected from harmful ones; to interact with harmful responses that are useful for their work \textit{without} being exposed to unhelpful, harmful responses; and to access the full scope of responses if they so choose.
To accomplish these, \SYSTEM introduces additional UI controls that provide the journalist greater agency in choosing when to encounter and interact with comments that are likely harmful. These controls are different depending on whether or not the likely-harmful comments might also be helpful to the journalist's work, so the journalist can more safely interact with potentially harmful yet helpful comments if they are compelled to do so. Below, we detail how \SYSTEM presents the content by walking through its different screens (Figure~\ref{fig:exploratory_design}) with a user scenario.

\textbf{User Scenario ---} Consider a journalist who recently tweeted about their coverage of a controversial issue, attracting a large volume of reader replies to their tweet that contains a mix of helpful and/or hateful content. The journalist can safely engage with those replies through \SYSTEM. First, assume the journalist wants to see readers' constructive and likely-harmless feedback. To do so, they can click the ``Show replies'' button on the tweet from the homepage, surfacing the replies that \SYSTEM has classified as non-toxic (Panel 2 in Figure~\ref{fig:exploratory_design}). To foreground the replies that are more likely to be helpful to the journalist's work, \SYSTEM also groups those classified as relevant at the top. The design of this screen is motivated by G1, as it provides a level of protection against the likely-harmful reader responses and helps the journalist to easily benefit from the utility provided by ``good'' comments (which might include sources, constructive critiques, and tips). We considered displaying the comments without grouping by relevance but decided against it, as this would not as effectively accomplish G1 given that relevant comments likely provide greater utility to journalists than irrelevant ones.

Next, assume the journalist wants to engage with critical feedback on their article. For example, they may want to understand what readers perceive to be gaps or bias in their coverage and anticipate that this feedback may be delivered in a heated way. To view these comments, they can click the ``Show hidden replies'' button, bringing them to the replies that \SYSTEM has classified as toxic but relevant (Panel 3 in Figure~\ref{fig:exploratory_design}). The design of this screen is motivated by G2, as it enables the journalist to interact with potentially harmful comments that could still be useful to them without needlessly being exposed to potentially harmful \textit{and} unhelpful comments. We considered displaying all of the toxic comments on this screen at once, with the irrelevant ones displayed under the relevant ones, but ultimately chose not to; we determined that this would not as effectively accomplish G2 given that the journalist might end up being exposed to irrelevant and toxic comments without wanting to, reducing their control over exposure to potential harassment.

Finally, assume the journalist wants to view the full scope of reactions to their article — including the comments that are unhelpful and toxic. For example, they may simply be curious, or they may want to be informed of potentially threatening replies. To do so, the journalist can click the ``Also show replies that are both irrelevant and toxic'' toggle (Panel 4 in Figure \ref{fig:exploratory_design}). This aligns with G3, as it preserves the journalist's ability to ``view everything.''

While iterating on the design of \SYSTEM, we also considered adding visual indicators to the screen shown in Panel 1 of Figure~\ref{fig:exploratory_design} to summarize the toxicity and relevance of the reader engagement for each tweet. We decided against this because it was unclear how we might summarize reader engagement accurately and meaningfully, especially in the context of how journalists reason about it in nuanced, granular ways. The exploration of designing effective feedback summarization in the presence of harmful content for various populations (e.g., journalists, politicians, professors, content creators) remains an important direction for future work.

\section{User Testing for \SYSTEM}
\label{sec:eval}
In this section, we present the results of a user study with eight journalist participants (P1-P8) who explored what audience engagement on social media would look like when mediated by \SYSTEM. Through user testing, we evaluate \SYSTEM's effectiveness at addressing the needs we synthesized from our need-finding interviews and discuss further insights from participant feedback.

\subsection{Study Design}
To evaluate \SYSTEM, we recruited journalists who were: (1) active on Twitter/X and (2) authored tweets promoting their own stories that other accounts replied to. We decided on these criteria because we developed our interface to integrate with Twitter/X and because an evaluation of our interface would only be pertinent if there were replies that might be considered toxic or relevant. Two of our participants were also participants in our need-finding interviews. We otherwise recruited six additional participants by cold e-mailing journalists that satisfied these criteria and asked participants for referrals to colleagues that were likely fits. Seven of our participants tested our system populated with their own tweet data, and one participant instead tested with data from a close colleague in the same newsroom and on the same beat.

The user testing was semi-structured with questions that guided participants through the different interface screens, where we encouraged them to talk through their thoughts and reactions while interacting with each page. The full set of the structured questions is available in Appendix~\ref{a:user_test_qs}. Because of the open-ended nature of these user tests, we asked additional questions when relevant. The user tests lasted between 20 to 45 minutes, varying based on the participants' past exposure to online harassment, their insights with protective tools and tactics, and their industry experience. We discuss the limitations of conducting a time-boxed user study in Section~\ref{ssec:limitations}. Participants were compensated for their time with a \$30 Amazon giftcard. Similar to the methods for the need-finding interviews (Section~\ref{sec:interviews}), two researchers coded the transcripts of the interviews using thematic analysis and an initial set of categorical labels informed by the user test questions that were expanded as needed. We again use IRR as a process to reduce confirmation bias and achieved a Kupper-Hafner agreement metric of 0.70, indicating high agreement. The two researchers discussed and resolved these conflicts to determine the final codes for the data. The full codebook can be found in Appendix~\ref{a:user_test_codebook}.

We do not report individual demographic information about our participants or disclose which newsrooms they are employed by because of the risk of deanonymization. However, the following are aggregated demographics information: three identified as men, and five identified as women; all of our participants are or were previously employed by major newsrooms, with the number of yearly subscribers to each newsroom ranging from hundreds of thousands to millions; the beats covered by our participants also varied --- four covered technology, two covered the environment, one covered politics, and one covered sports. 

\subsubsection{Ethical Considerations}
Because of the sensitive nature of this research, we informed all of our participants about the goal of this study before we began the interviews. Although we presented all but one of our participants with content that they had actually received on their Twitter/X accounts, we made sure to inform all participants of the potentially triggering nature of the content they were about to consume in our study. Additionally, we reminded participants that they could decline to answer any questions and were able to stop the interview at any time, and we obtained their consent to record the interviews. Because of both the online and offline threats to journalists' safety, we treat the information our participants shared with us as highly sensitive. As such, we again remove PII from any interview excerpts included in the paper and only present aggregate descriptive statistics to protect participant anonymity. This follow-up work received approval from our IRB.

\subsection{Findings}
In this section, we present the findings from a series of user testing interviews of \SYSTEM with journalists who experienced significant harassment on Twitter/X. Through this evaluation, we find that participants: 
\begin{enumerate}
    \item felt effectively protected by \SYSTEM against harassment and believed it could generalize to serve other visible and vulnerable groups,
    \item valued customization in automated tools,
    \item expected and made sense of errors in automated classification,
    \item did not feel valuable reader engagement was hindered by \SYSTEM,
    \item were keen of \SYSTEM's application in other contexts, and
    \item surfaced a need for a tool to discern physical threat from other harassment.
\end{enumerate}

\subsubsection{Participants felt that \SYSTEM effectively protected them against harassment and could generalize to serve other online users in the public eye.}
Throughout the interviews, it was apparent that preferences for dealing with online harassment varied --- even within a curated set of journalists who had previously dealt with online harassment. Still, all participants agreed that the option to use a system like PressProtect would likely benefit many journalists. Four participants, including P8, added that this tool would be especially relevant to the experiences of marginalized-identity journalists or those that report on sensitive beats:

\myquote{``I think journalists of color [would benefit from this tool], as well as female reporters---anyone in a marginalized community would get more harassment and more attacks than an average white dude\ldots also, reporters covering issues like immigration, gender, racial justice, investigative reporters. Those are the groups of people who may benefit more from this tool than other reporters.'' -- P8}
    
\noindent When examining what aspects of \SYSTEM~participants appreciated, we found that participants saw value in the nuance of its underlying logic, which considers the relevance and harmfulness of comments without binarizing them as ``good'' or ``bad'':

\myquote{``[PressProtect] reminds me of [Block Party], the way it's trying to filter responses and give you a little bit more control over what you're looking at, so I think this could be useful.'' --~P1}

\myquote{``I really liked the fact that [PressProtect] separates [comments] into three groups instead of two. When you look at Twitter and some of their implicit tools, even ones that worked a little better a couple months ago before the takeover, to look at [what Twitter labeled as] harmful replies, it's just based on language. They are all categorized as, `these are okay', and `these are not,' and I think [the different logic] is what is good about [PressProtect].'' --~P4}

\noindent Similarly, participants enjoyed the ability to filter comments that were unlikely to lead to meaningful discussions using relevance. For instance, P5 explained how the relevance axis of \SYSTEM's abstraction was actually the one they might personally find the most utility in:

\myquote{``If I look at all the [tweets] that were hidden, they're not tweets that I would have, under almost any circumstances, responded to---not necessarily just because they were harmful---but also because they weren't helpful or real conversation topics\ldots I think there are people who would emphasize different parts of this tool. For me, I think it's more about filtering out irrelevant stuff.'' -- P5}

\noindent More broadly, four of the participants felt that \SYSTEM's underlying logic could also serve others that are required to maintain an online persona in the public eye, such as celebrities, politicians, or social media influencers. 

\subsubsection{Participants value the ability to customize automated tools for a wide range of personal preferences.}

To understand the extent to which \SYSTEM's axes of toxicity and relevance could represent individual journalist's preferences, we investigated how each journalist defined these classifications. When asked to define what ``toxicity'' meant in the context of professional online engagement, our participants unsurprisingly gave various definitions. Some described toxic interactions as ones that involve insult, abuse, or identity-based harm, and others perceived toxic interactions as ones that were inherently irrelevant, describing them as as ones that did not ``engage substantively with the story.'' Others discussed how the scale of engagement could also make otherwise manageable interactions feel like harassment. Seven of our participants acknowledged that there are many factors that shape their personal thresholds for engaging online, such as past experiences of harassment, newsroom-dependent standards, and the various threat models they are vulnerable to (e.g., nation-state censorship, personal stalker).
    For instance, P7 described the variable degree of firsthand triaging journalists might have to undertake depending on how well-resourced their newsrooms were:
    
    \myquote{``If you're in a newsroom that provides a lot of support, or doesn't provide a lot of support, that definitely shapes the way that you would experience [harassment]. [My newsroom] provided a lot of support\ldots Other newsrooms that I worked at wouldn't provide anything, so there would be a much bigger emotional burden of needing to deal with it myself, needing to figure it out and triage what I'm receiving.'' -- P7}
    Consistent with intersectional feminist theory, four of our participants engaged with how various identity categories (e.g., gender, sexuality, race) shape harassment experiences, and as a result, their preferences for using social media.
    One participant explained their personal level of comfort with seeing most of the comments they receive comes from the degree of harm from identity attacks they anticipated as a cis male. They noted that other marginalized-identity journalists might have a different threshold for what content they are willing to expose themselves to by default:
    
    \myquote{``I want to be very clear: I am a cis male writer, so certainly I'm getting race-based attacks, but it's different as a woman in the industry; it's different for somebody who's queer in the industry.''}
    Other facets described by our participants that influenced their thresholds included what past harassment experiences they had (e.g., if they had a story go viral, if they had been a target of nation-state censorship), or if their beat was particularly sensitive. The expectation that some areas of reporting attract more harassment than others was well-known to participants, and P7 elaborated on how this made such beats self-selecting for those with ``thick skin'':

    \myquote{``For journalists who are on politically charged beats, like China, the alt-right, or Trump---these are beats that are known to attract a lot more harassment. There are some job posts that actually indicate that you need to have a thick skin because of the level of harassment the journalist receives. So, in a way\ldots the more contentious beats actually self-select for people that can just deal with it. I wouldn't say that everyone can.'' -- P7}

\noindent
Although participants felt that the axes of toxicity and relevance aptly characterized their engagement preferences for a given comment, the many facets of each journalist's experience and identity shaped their individual thresholds for toxicity and relevance. Five of our participants felt that \SYSTEM~would better serve a wider range of journalists if they could customize these feature thresholds. For instance, two journalists suggested optimizing \SYSTEM~for a low false positive rate when on the job; three others proposed adding a finetuning mechanism for \SYSTEM's classifications of toxicity and relevance to better align with their individual definitions. P1 elaborates on how they imagine finetuning an automated tool might integrate into their workflow:

    \myquote{``I think there's definitely some sort of finetuning that could be done with the filtering. I think it could be potentially useful to allow the person using the tool to change the classification, if they feel like it's incorrect. In the example that we talked about earlier, I was like, this person is being very polite in their tweet, but there's a potential for harassment. It would be nice to be able to toggle that to your toxic replies and save that in a separate folder, so you can return to those accounts and see what they're doing. Or, if you see a tweet that's misclassified as toxic, and you're like, Oh, this is actually a relevant reply, I want to move it back to my main replies, you could click into these little icons and change how they're [classified]. I think that could be helpful.''}

\subsubsection{Participants expected and rationalized errors in automated classifications.}
    All participants acknowledged that any automated classification tool is bound to make decisions that deviate from their own. Seven participants described how PressProtect performed as well as or better than their expectations for an automated tool. Some categories of content that P5 perceived as out of scope or difficult for the system included jokes or sarcasm, which they also felt real people might struggle to discern:
    
    \myquote{``I don't have a huge issue with the way that [replies are] categorized --- even the joke tweets, but that would probably be a little bit of a blind spot [for the tool]. And again, frankly, that's a blind spot for real people on the internet trying to figure out what's going on too, so I certainly don't expect any AI model to be able to 100\% [categorize comments correctly].'' -- P5}

\noindent Another participant felt that it was unreasonable to expect any automated tool to incorporate relevant context or a body of external knowledge about the story in its classifications:

    \myquote{``This one maybe is misclassified\ldots I think this would be a really tricky thing to do automatically, because if someone is referencing a known public figure that's involved in that story somehow but not discussed in the article, I don't know how you would sort that or classify that without reading all the stories about this and knowing the context. So that makes sense to me, why it's classified the way it is.'' -- P2}

\noindent
Still, the consensus among five of our participants was that these disagreements between the classifications and their own perceptions were an acceptable cost when compared to \SYSTEM's benefits. For example, P4 imagined a workflow using \SYSTEM as a first-pass filter to decrease their manual triaging workload, a process they were already accustomed to with e-mail spam: 
    
    \myquote{``I feel like [the risk of false positives] is a fair trade off for me\ldots I'd be fine with having to take a little more time to sift through the hidden replies. At work, we have a tool that's really similar to this, but for e-mail spam [where] most e-mails will come through, and then a couple e-mails will go to quarantine.'' -- P4}

\subsubsection{Participants did not feel PressProtect would hinder valuable interactions with readers.}
    Although in our need-finding interviews, we found that it was crucial that journalists be perceived as receptive to reader feedback, consistent with prior work ~\cite{lewis2020online}, all of our participants expressed a cynicism for the quality or value of engagement they were likely to receive on social media. 
    Three participants specifically expressed distrust in the quality of interactions on the platform after Elon Musk's acquisition, citing upticks in spam and abuse and the shutdown~\cite{blockparty2023hiatus,block-together} of third-party tools that had previously helped them navigate Twitter/X. P1, especially after Musk's Twitter acquisition, adopted a personal policy to not engage on Twitter/X, and they felt that the professional value of Twitter/X replies had plummeted:

\myquote{``I think there's such limited value in a Twitter reply---even non-toxic and technically relevant Twitter replies for the most part---that anything that you would lose [by incorrectly categorizing content], you're going to be losing something that would have been of only very marginal benefit anyway.'' -- P1}

\noindent    Additionally, P7 felt that even missing reader responses that were not harmful was not a critical issue for a protective tool:

    \myquote{``I'm not really worried about missing good comments. I receive so many inbound things every day, whether it's a Twitter comment or LinkedIn message or an email, and my policy is I just get to what I get to. I'm not going to kill myself trying to reply to every single one, because it would just be way too much time, and I wouldn't get any work done. So things like good comments that I don't see and don't get to engage with, I probably wouldn't have engaged with it anyway.'' -- P7}
    
\noindent    As such, we observe that participants' views of engagement integrity and quality were closely tied with their trust in the platform and the platform's ability to effectively moderate content. While six participants acknowledged the drawbacks to mistakenly categorizing comments as toxic (i.e., hampering audience engagement in participatory journalism, missing a connection to a potential source), these participants were not deeply concerned that it would negatively impact their already unsatisfactory journalist-audience experience on Twitter/X.

\subsubsection{Participants were keen on the application of PressProtect's features in other contexts.}
    Several of the participants noted that the underlying logic of PressProtect could be leveraged or extended in other ways. Two participants specifically mentioned the potential of using this automated categorization mechanic as a triaging tool to help decrease the manual burden of safety teams in newsrooms. For instance, P7 explained how incorporating PressProtect into a triaging pipeline could allow journalists to assess risk  without direct exposure to the content: 
    \myquote{``I think if a newsroom had a protocol where they took responsibility and ownership over monitoring things for you, then there could be a really great benefit where [PressProtect] directs all of [the harmful comments] to the newsroom security team. So someone is seeing it, but not you. And then they can tell you what their risk assessment is without you having like been exposed to like the very direct toxicity.'' -- P7}



\noindent The proposed use of \SYSTEM as a triaging tool mirrors the comment moderation pipeline explored by Moderator, which uses classifiers to group similar comments for easier review by human moderators~\cite{jigsaw2016moderator}. Some other the potential ways to customize the tool discussed by our participants included the ability to do the following: share or offload the burden of triaging replies with another trusted entity (i.e., a colleague, similar to friendsourcing), toggle between sorting comments by relevance or in chronological order, or toggle the view of comment categories depending on if they are using Twitter/X professionally in that moment or are anticipating a flood of comments.

Participants also noted that PressProtect could serve journalists that do not have the resources and support of a dedicated newsroom security team, echoing the reflections of J3 on the need to democratize online harassment protection via third-party tools (Section~\ref{ssec:needfinding_results}). P1 and P4 described that PressProtect's logic could also benefit journalists when applied to e-mail and direct messages. While P4 noted that their newsroom provided filtering tools for e-mail, the filter's purpose was to protect against spam and scam, rather than harassment. In addition, P8 suggested that \SYSTEM could be extended by integrating with platform moderation mechanisms, such as reporting and blocking, to streamline protections against state-sponsored harassment, which often leverages bot accounts.


\subsubsection{Participants wanted a distinction between imminent, physical threat and other forms of online harassment.}
    As discussed above, participants expected and were generally not concerned with the ramifications of falsely labeling non-toxic and even relevant tweets, or false positives. However, a theme that emerged in five interviews was the importance of flagging urgent threats to their physical safety and not losing this signal by filtering such comments from default view or incorrectly classifying them as non-toxic. One participant was not concerned with ``losing edge cases'' for ad hominem verbal attacks, but they were concerned with (1) losing contact with a colleague or potential source or (2) not seeing a threat that could potentially escalate. Still, P1 felt that the tradeoff between the frequency of these events and the tool's utility might be worthwhile:

    \myquote{``I think both of those scenarios [1 and 2] are pretty unlikely, so it's a level of risk that would seem worth the benefit of having a lot of the most obnoxious stuff filtered out.'' -- P1}

\noindent
    Others were less compromising on flagging such threats and the possibility of filtering them from view. Although P7 acknowledged that PressProtect could likely serve many journalists, P7's frequent participation in public speaking events coupled with the smaller scale of engagement they receive on social media shaped their preferences for monitoring their online accounts; as a result, P7 preferred to filter \textit{nothing} from their default comment view because identifying physical world threats took utmost priority over convenience and usability:
    
    \myquote{``I would just stop using [PressProtect] because of my bias towards wanting to see everything. I don't necessarily care what order I see [comments]. I just want to, at some point, have seen them all\ldots You still really need to monitor what people are saying to make sure that doesn't translate into physical danger.'' -- P7}


    Two participants added that newsrooms struggle to combat online threat vectors to journalists' safety. P7 explained how protecting journalists from online harassment has been a ``huge weakness across the industry,'' stating:

    \myquote{``There isn't really a standardization or understanding of how to provide basic protections against digital harassment. There's much more of an understanding of physical harassment, like getting you a bodyguard.''}

\noindent
    However, to benefit from physical protections like a bodyguard, journalists must first be equipped to identify when their physical safety is threatened. This highlights a gap in the resources for assessing risk that must be addressed for journalists to take the necessary steps to obtain physical protections.
    Similarly, P8 detailed their experience with the failings of organizational and platform support when facing a state-sponsored online harassment campaigns:

    \myquote{``I just don't want to engage with people on Twitter, period. But I think for me, it's different because I'm more traumatized by state-sponsored harassment because that could lead to [feeling] physically unsafe. I feel violated and harassed when I see the state-sponsored [online] violence.''}
    
    \myquote{``In my organization, they provide digital security security training [as a prevention tactic for online harassment], which I don't think is very helpful, because I'm like, I know more than you do. And my situation is too unique for you to be able to give me a recommendation.''}

    \myquote{``I had to go to our security people here at [my newsroom] and ask them [to report state-sponsored bots] because they know people at Twitter, and then they contacted whatever department at Twitter directly to ask them to monitor and remove the bots. I reported many times myself: nothing worked.''}




\noindent    P8's experience underscored that a sense of online and offline safety can be deeply intertwined, and that in some cases, the harm from harassment is not simply contained to the message content. Rather, these campaigns publicly signal that these journalists are being monitored, causing them to self-censor and fear for their physical safety. Yet, both their newsroom and Twitter/X as a platform have not served their needs to identify and shut down these attacks. Additionally, the only avenues that were effective for P8 --- (1) taking down state-sponsored accounts through a backchannel between their newsroom and employees at Twitter, and (2) self-censoring their online posts --- are not necessarily desirable or accessible solutions for all journalists. This is especially true in the face of major changes in journalists' attitudes toward and trust in the platform. We note that the ability to identify these dangers is not currently accounted for by PressProtect because it builds on the Perspective API's toxicity clssifier, which does not disambiguate different kinds of harmful content, such as offensive language versus physical threat. We discuss this gap and its implications in more detail in Section \ref{sec:discussion}.


\section{Discussion}
\label{sec:discussion}
In this paper, we present the following contributions: (1) we formalized journalists' unmet online safety needs, and (2) we developed an effective abstraction for journalists interfacing with social media in the face of significant harassment, using the axes of comment toxicity and comment relevance to their work. We discuss the implications of these findings on the design of anti-harassment tools and the roles of the stakeholders in the online journalism ecosystem.

\subsection{Designing for Visible and Vulnerable Populations}
To effectively design harassment protections for vulnerable communities, we must first make sense of the various \textit{dimensions} of a community's needs and how the \textit{nuances} of these dimensions compare to those of other communities\. 
In Sections~\ref{sec:interviews} and~\ref{sec:eval}, we investigated how journalists experience, process, and manage social media engagement in the face of significant harassment. Informed by our findings, we identify the following dimensions of journalists' needs regarding online engagement and discuss the nuances of each:

    \dimension{Utility}:
    The utility that journalists derive from online engagement (e.g., promoting their work, connecting to sources and their professional network, understanding reader responses) necessitate that journalists be active and visible on social media. Further, journalists want to view the work of their peers and readership opinions to inform their work. This centrality of visibility dictates that journalists may experience fewer benefits when restricting the visibility of their audience or their own accounts through the primary safety mechanisms offered by platforms (e.g., blocking or muting accounts).
    
    \dimension{External Constraints}:
    Both the public and journalists' own perceptions of journalistic professionalism reduce options for effective harassment mitigation, as journalists feel the need to \textit{appear} open to criticism; we found that if journalists fail to maintain this public-facing image, they fear triggering accusations of being biased or sloppy, or pushing a conspiratorial political narrative. Upholding this receptive appearance does not require journalists actively reply to readers, but it does make existing moderation levers, like blocking accounts, unsuited for their professional norms. Furthermore, journalists' usage of social media platforms is subject to news organizations' explicit and implicit social media policies that encourage their active engagement in online spaces.
    While these policies can be ambiguous, violating them has cost journalists their jobs~\cite{lee2023journalists}.
    Because of these external constraints imposed by the public and their employers, journalists must cultivate ``authentic''~\cite{nelson2021tightrope} journalist-audience relationships in online spaces through open and active participation. While other populations, such as content creators or politicians, face similar reputational risks on social media, journalists must uniquely consider direct risks to their employment. 
    
    \dimension{Safety}: 
    Journalists' safety needs are shaped by the specific nuances of visibility and compulsory participation online. While platforms and some newsrooms provide resources for journalists facing imminent danger, the needs to (1) identify and document these threats and (2) monitor abusive patterns that may escalate to such threats fall upon journalists themselves to manage. Therefore, they must engage with online comments, even if they are harmful or triggering, to triage and escalate as necessary for their physical safety. Existing platform moderation mechanisms (e.g., blocking users, removing offensive content) binarize content as ``benign'' enough to be viewed or ``harmful'' enough to be filtered out completely; however, the nuances of journalists' safety needs make these approaches impracticable and necessitate that platforms \textit{mediate} journalists' navigation of potentially harmful content in a safer way, rather than dispose of such content altogether. 
    
    \dimension{Empowerment}:
    Despite their intended purpose, platform abuse prevention levers such as blocking are not aligned with the journalists' sense of agency and empowerment. Because platforms like Twitter/X indicate to users when they have been blocked, acknowledging and addressing harassment through blocking can instead empower abusers as positive feedback of successful harassment attempts. Other common strategies, such as disengaging or self-censoring, similarly disempower journalists and are at odds with the principle of press freedom. We observed this with one of our participants, who felt that a nation-state had succeeded in its attempt to silence them: ``The intention [of the attack] is for me to stop doing what I do now and to be silenced precisely. And I think in some way, they're successful because I'm no longer active on Twitter. I'm just like a bot promoting stories'' (P8). As such, the nuances of journalists' needs regarding empowerment in the face of online harassment have made existing strategies both ineffective and undesirable.

Sarah Jeong, a journalist whose social media controversy and subsequent harassment led her to quit her editorial position at The New York Times, encapsulates the interplay of these different dimensions in her experience with this flood of online attacks~\cite{jeong2023goodbye}: 
    \myquote{``Twitter offers two tools to theoretically protect yourself, [blocking and muting]. Since the platform indicates when you’ve been blocked by a user, the Times asked me not to do it to anyone. Besides, the most motivated of my haters would make new accounts anyway. Large, influential accounts could drum up fresh troops to send forth into my replies\ldots I wasn’t sure what was more unsettling: getting a death threat and seeing it, or not seeing~it.''}

Jeong's experience was shaped by the nuances of the external constraints (from The Times, despite its past claim to ``support the right of [their] journalists to mute or block people on social media who are threatening or abusive''~\cite{nyt2017guidelines}) and safety needs she had as a journalist, which ultimately made existing protective tools nonviable.
Through this systematization of journalists' needs, we argue that framing needs as multi-dimensional offers a lens for reasoning about the needs of other populations, particularly those that are also both visible and face outsized harassment online. Therefore, by considering the various dimensions of different communities' needs, we can better design platform affordances and protections that actually address the needs of vulnerable groups.
Furthermore, we argue for using this framing comparatively for groups. For instance, to what degree can \SYSTEM's abstractions benefit groups such as social media influencers or politicians? How granular or broad are the groups that can share such abstractions? How similar are the needs between all visible and vulnerable groups? We argue that understanding different group needs in this manner can guide more thoughtful platform design for a diverse userbase. 

Beyond a harm-centric research approach to supporting marginalized communities, this multi-dimensional framing may also be applied to the desires of such communities that will enable them to flourish, rather than simply survive. Recent work in the interactive design space has highlighted the importance of fulfilling the desires of marginalized communities, particularly Black, Indigenous, and People of Color (BIPOC), to find joy, affirmation, and liberation in online spaces~\cite{to2023flourishing}.
Understanding the nuances of these desires (e.g., belongingness as a nuance of a desire to be empowered, cultural heritage as a nuance of a desire to express oneself) can likewise better inform designs that allow marginalized communities to flourish in online spaces.
    
\subsection{Roles of Stakeholders}

\subsubsection{Newsrooms}
Prior work in the computer security community has assumed that journalists (1) have viable alternatives to seeking audience engagement and (2) are effectively supported by their organizations to stay safe online ~\cite{samermit2023millions}. However, our findings (Section~\ref{sec:eval}) indicate that journalists are often subject to newsrooms' social media policies that require their participation, yet do not feel that these newsrooms adequately support them against online harassment.
Our participants discerned that protections offered by newsrooms focus on addressing physical rather than online harassment. To effectively leverage these protections, however, journalists themselves must identify and document when threats become physical, manufacturing a safety need for journalists --- or in the case of well-resourced newsrooms, their safety teams --- to be able to monitor and triage high volumes of engagement for dangerous behaviors. We believe that the use of \SYSTEM's abstraction to make sense of comments' harmfulness and helpfulness, coupled with the development of automated tools to identify physical threats, can help journalists and newsroom safety teams to more effectively make sense of large-scale harassment.

A more fundamental tension underlying journalists' experience of online harassment comes from an industry shift toward social media as the primary avenue for reciprocal journalism~\cite{lewis2014reciprocal}. The literature in journalism practice defines reciprocal journalism as a concept that positions journalists as community organizers to build trust and social capital with their audience. However, the theory of reciprocity requires mutual ``gift-giving'' which, in this context, manifests as valuable thoughts, perceptions, or behaviors~\cite{lewis2014reciprocal}. We observed that journalists lost faith in Twitter/X's capabilities to: (1) enact effective platform moderation and (2) foster a userbase with ``gifts'' to give. We argue that, as it stands today, Twitter/X's ecosystem has thereby made reciprocal journalism nonviable in practice for many journalists. In light of existing platform and organization limitations, we recommend that newsrooms reconsider both the explicit and implicit expectations that are imposed on their staff to better center the needs of their most vulnerable journalists --- who are often those of marginalized identities --- paralleling design recommendations made by the CSCW community~\cite{ashktorab2016designing,schoenebeck2023online,scheuerman2018safespaces,Blackwell2017classification,han2023hate}.

\subsubsection{Social Media Platforms}
Twitter/X has positioned itself as a platform where ``the public has access to reliable information from trusted sources,'' even publishing a guide to ``how journalists can use Twitter safely'' with suggestions to use its native tools for addressing account security and online harassment threats~\cite{twitterjournalists}.
However, both prior work~\cite{unesco2021trends} and our data indicate that journalists view social media platforms as the primary channels for and enablers of online violence; moreover, the tenuous protections offered by existing platform safety tools have made journalists continue to feel unsafe.
We observed how journalists have lost trust in the platform, especially after Musk's acquisition of Twitter, to counter abuse despite its proclaimed efforts, as well as in the platform's users to engage in high-quality interactions. As Twitter/X has become an inhospitable ecosystem for reciprocal journalism, we argue that if platforms wish to re-position themselves as central, usable, and useful to journalists, they must address this issue of lost trust.
Prior work in information systems management has identified benevolence trust (i.e., trust that a platform will act in goodwill rather than opportunistically) as the most significant factor in users' continuation on a social media platform~\cite{wu2013benevolence}. 
Drawing from this,
we argue more broadly that designing for vulnerable users can help re-build trust with those that have experienced harm and thereby aligns with platforms' incentives; it behooves platforms to prioritize building and maintaining trust with its diverse user groups to sustain their platform usage.

Beyond the design of platform affordances and safety measures, we highlight the unrealized potential for third-party developers in this ecosystem. Past CSCW work has discussed how community or third-party developers are uniquely positioned to rapidly build platform tools to address vulnerable communities' needs because they are unhindered by the frictions that platforms face when developing features or policies~\cite{han2023hate}. While most of our participants did not use third-party tools to manage harassment, several of our participants mentioned using or knowing about Block Party. Block Party, however, is one of many third-party Twitter/X applications that shut down due to Musk's imposition of prohibitive API prices and degraded API access~\cite{binder2023twitterapi}. 
Between Twitter/X and Reddit, this trend of stymieing developer access to platforms is a threat to the work of both third-party developers and researchers~\cite{calma2023twitter}.
Although Twitter/X has developed safety features (e.g., blocking and muting accounts, muting notifications), we demonstrate how these fail to meet the needs of journalists, and likely those of other visible and vulnerable populations. We argue that instead embracing the distinct role of external developers in platform and community governance will: (1) rebuild trust between platforms and their users, which will (2) enable reciprocal social media exchanges, and (3) democratize access to effective community safety tools. We situate the combined strengths of centralized (platform) and decentralized (end user, third-party developers, newsrooms) actors as a case that multi-level governance systems~\cite{jhaver2023decentralizing} may better suit governance in online social media platforms that have a diverse user population with vulnerable groups.

\subsection{Limitations}
\label{ssec:limitations}
Journalists are not a monolithic group: even within both of our studies, each with eight participants, we observed a large range of personal experiences, identities, and preferences for engaging online. Thus, the approach of finitely distilling the needs of a group, such as journalists, is inherently limited and cannot address all of the nuanced needs of every member. 

In addition, both our need-finding and user testing studies are limited by the scope and scale of our recruitment processes. In this work, we began the participatory design process with AAPI journalists that self-identified as having experienced online harassment and self-selected to participate in our study. Therefore, we do not claim that our participant samples are representative of all journalists. Still, consistent with Blackwell et al.'s conclusion that centering the needs of marginalized groups will produce a design process that better serves all users, we found that our need-finding process with AAPI journalists yielded a set of design goals that, when actualized in \SYSTEM's implementation, effectively protected a broader population of journalists, as our user testing participants were not recruited from a specific identity group. During user testing, we note that one journalist surfaced the additional need to discern physical threats from other kinds of harassment. Although participants in our initial need-finding study discussed experiences where online harassment had escalated to physical threats and the impacts of these experiences, the need to identify physical threats was not clearly articulated by any participant. This may be because of the open-ended nature of our semi-structured interview questions, which did not prompt participants specifically about online safety needs; on the other hand, with the structure and context provided by our card sorting exercise of inflammatory tweet mentions, the restrictiveness of closed card sorting may not have allowed participants to discuss alternative ways they would have liked to process harmful content.

We conducted our user testing study as a time-boxed exploration of \SYSTEM's interface with a small, curated set of participants' past tweet data. This approach is limited in that it fails simulate journalists' in situ usage of \SYSTEM as a long-term solution. When we set out to build \SYSTEM in February~2023, we envisioned deploying it as a real-world tool that integrated with Twitter/X through its API, hooking into the notification system of the platform itself as the mechanism for controlling the journalist's exposure to reader engagement. However, as access to the Twitter/X API degraded and became prohibitively expensive just weeks later~\cite{twitter2023api}, 
\SYSTEM's deployment as a usable application became impossible. As a result, our goal shifted to understanding the effectiveness of \SYSTEM's underlying abstraction to reason about content in a way that is helpful to journalists, rather than to evaluate a real-world system's performance or usability in practice. We hope that the insights from our design exploration with \SYSTEM can be integrated into future platform or tool designs. 

\section{Conclusion}
\label{sec:conclusion}

In this work, we formalized the outstanding needs of journalists to safely participate in online spaces when facing significant harassment, drawing from the findings of our eight need-finding interviews. We synthesized a set of design goals to make journalists' social media experience safer and more usable, from which we built \SYSTEM, an interface for Twitter/X that leverages a logical abstraction for reader comments and uses UI controls for viewing comments as its protective mechanism. Through user testing with a separate set of eight journalists, we demonstrated that \SYSTEM's abstraction and protective mechanism effectively served journalists; however, user testing also revealed a previously unidentified need for journalists to be able to discern physical threats from other harassment. Our findings deepen our understanding of vulnerable communities' needs as multi-dimensional and central to safe, effective platform design and suggest opportunities for collaboration between news organizations, platforms, and third-party developers to better achieve multi-level governance.

\section{Acknowledgements}
\label{sec:acks}

We thank our participants for their time and openness in sharing their experiences with us and providing formative feedback to refine the design of \SYSTEM. We are also grateful to Waliya Lari and AAJA for connecting us to their community. We thank Michael Bernstein for his insight on the framing and design of our need-finding study and Marissa Henri for her help designing \SYSTEM's user interface. This work was supported in part by a Magic Grant from the Brown Institute for Media Innovation, a Sloan Research Fellowship, an NSF Graduate Research Fellowship \#DGE-1656518, and NSF grants \#2030859 and \#2127309 to the Computing Research Association for the CIFellows Project. 

\bibliographystyle{ACM-Reference-Format}
\bibliography{paper}

\received{January 2024}
\received[revised]{April 2024}
\received[accepted]{May 2024}

\appendix
\label{appendix}

\section{Need-finding Interview Questions}
\label{a:needfinding_qs}
\subsection{Questions about demographics and Internet usage}
\begin{itemize}
    \item How long have you been a journalist?
    \item What types of stories do you usually cover?
    \item What social media platforms do you regularly use? 
    \item What do you primarily use social media for?
    \item Is social media usage a core part of your professional career? 
    \item What social media platforms are most important to you and your professional career?
\end{itemize}

\subsection{Questions about harassment experiences, harms, and defenses}
\begin{itemize}
    \item Can you tell me about a time when you experienced online hate and harassment on Twitter?
    \item What was the experience like? What happened, and how was it triggered?
    \item What were the types of messages that were sent to you? How were they sent – via a DM? A mention?
    \item Can you walk me through an example of unwanted hate and harassment you may have received online recently? Feel free to show me the concrete example on the social media platform.
    \item What did you do in response?
    \item To what extent do you feel like your identities (e.g., gender, occupation) played a role in the abuse you received? 
    \item Outside of your identities, are there other aspects that may shape the harassment you experience?
    \item What resources or tools do you use to stay safe from online harassment?
    \item Are there any strategies or tools you’ve tried that you didn’t find effective?	
    \item Do you ever block or mute accounts on Twitter?
    \item How frequently do you do so?
    \item Have you ever used any other tools, or are you aware of other tools that either you or your colleagues use to stay safe from online harassment? 
    \item How do you learn about new resources?
    \item What factors into whether or not you try these resources yourself?
\end{itemize}

\subsection{Open-ended questions}
\begin{itemize}
    \item Is there anything we didn’t talk about that you would like to share?
\end{itemize}

\section{Perspective API Threshold Experiment}
\label{sec:threshold}
\begin{table}[h]
    \centering
    \begin{tabularx}{0.7\columnwidth}{Xrrrr}
        \toprule
        Model @ Threshold   &   Accuracy    &   Precision   &   Recall  &   F1 \\
        \midrule
        TOXICITY @ 0.5      &   70\%        &   0.58        &   0.66    &   0.62 \\
        TOXICITY @ 0.7      &   70\%        &   0.68        &   0.32    &   0.44 \\
        TOXICITY @ 0.9      &   65\%        &   0.89        &   0.06    &   0.13 \\
        \midrule
        SEVERE\_TOXICITY @ 0.5  &   64\%    &   0.93        &   0.04    &   0.06 \\
        SEVERE\_TOXICITY @ 0.7  &   63\%    &   1.0         &   0.0     &   0.0 \\
        \bottomrule
    \end{tabularx}
    \caption{\textbf{Perspective API Scores for Twitter Content}---%
        We show the performance of the Perspective API across several thresholds and classifiers for a hand-labeled dataset of comments from Twitter. We utilize the TOXICITY classifier at a threshold of 0.5 for our study as it best balances precision and recall.
    }
    \label{table:twitter_scores}
\end{table}

To identify an appropriate Perspective API threshold for \SYSTEM, we leverage a dataset of toxicity annotations curated by Kumar et.~al~\cite{kumar2021designing} which provided toxicity labels of 107K~comments sourced from Reddit, Twitter, and 4chan. Each comment was labeled by five~raters on 5-point Likert scale from ``not at all toxic'' to ``very toxic.''. We collapse this score into a binary toxicity rating based on if the median toxicity rating was greater than ``moderately toxic.'' We build a balanced dataset of 5K~comments sourced from Twitter and evaluate how well the Perspective API's TOXICITY and SEVERE\_TOXICITY classifiers perform, shown in Table~\ref{table:twitter_scores}. We observe that the Perspective TOXICITY model best balances precision and recall (F1 of 0.62, aligned with prior work~\cite{saveski2021structure}), and so we utilize this threshold for the design of \SYSTEM.

\section{User Test Questions}
\label{a:user_test_qs}
\begin{itemize}
    \item Here we have a curated view of some Tweets you’ve written in the past and filtered their replies. Can you talk us through your initial observations and reactions to what you see here?
    
    \item Is there anything confusing or ambiguous about the elements on the page?

    \item Here we have a page view that displays a Tweet you wrote with the replies beneath it. Can you walk us through your thoughts as you view the elements on the page? 
    
    \item You might have noticed that there are two versions we have prepared at the top of the page. Version A groups the nontoxic relevant replies at the top, and the irrelevant ones at the bottom, but if you select Version B, it has all of the “nontoxic” replies listed in chronological order. Which of these do you prefer?
     
    \item For each of these versions, what benefits or drawbacks do you anticipate? 
    
    \item Now, if you click the “Show hidden replies” button, you’ll see a dialog.
    
    \item (If you’re comfortable with this, can you view the quarantined content?)
    
    \item Why do you think these comments were filtered?
    
    \item Do you agree with the decision to filter these comments? Why or why not?
    
    \item Why do you think these comments were filtered into this section?
    
    \item Do you agree with the decision to filter these comments? Why or why not?
    
    \item Overall, what type of content do you think this filter fails to catch, if any?
    
    \item What benefits and drawbacks do you see with filtering content like this?
    
    \item How would a tool like this be integrated into your existing workflow on Twitter?
    
    \item What kind of content do you interact with regularly that would and would not be addressed by this tool?
    
    \item What were you hoping for (or expect) but did not get from this tool?
    
    \item Can you describe what your mental model is for reasoning about how toxic or relevant a comment is?
    
    \item What are your expectations about how well an automated tool can simulate decision-making for deciding what is toxic/relevant?
    
    \item What components did you find confusing or difficult to navigate?
    
    \item What populations would be served with this tool? Are there any populations that this tool would not benefit? What about for different areas of reporting?
\end{itemize}

\section{User Test Codebook}
\small
\begin{longtable}[h]{p{0.3\linewidth}p{0.6\linewidth}}
\toprule
Code                                        &       Meaning \\
\midrule
\SYSTEM logic (pro/con)                         &       Discussion of \SYSTEM's toxicity-relevance logic as useful or unhelpful \\
\SYSTEM sorting                             &       Discussion of how the prioritization of relevance impacts their experience with \mbox{Twitter/X} \\                   
UI control mechanism (pro/com)                  &       Discussion of how filtering in general can be useful/undesirable to them \\     
Use case for journalists                    &       Discussion of (certain groups of) journalists (e.g. identity, beat) that might benefit from \SYSTEM \\               
Use case for other populations              &       Discussion of other groups of people that might benefit from \SYSTEM \\                   
Forgiving classifier errors                 &       Comments on how the participant normalizes or forgives ``mistakes'' made by \SYSTEM or, more broadly, automated systems \\               
Disagreement with \SYSTEM                   &       Comments on how they do not agree with the tool's classification \\               
Explanation for disagreement                &       Comments on why they do not agree with the tool's classification \\               
Error cost (pos/neg)                       &       Comments on the cost of errors that are net positive or negative \\ 
\SYSTEM performance                         &       Reflections on how \SYSTEM performed relative to expectations \\       
\SYSTEM error expectation                   &       Comments on expecting \SYSTEM to make errors \\               
Interaction quality (pos/neg)              &       Positive/negative comments on quality of interactions they have on \mbox{Twitter/X} \\       
Ecosystem quality (pos/neg)                &       Positive/negative comments on the \mbox{Twitter/X} ecosystem and any anti-abuse mechanisms \\ 
Platform trust                              &       How much trust the participant or their colleagues place in \mbox{Twitter/X} as a platform to moderate its users and the content they produce fairly \\   
Twitter acquisition                         &       Comments specifically related to Elon's acquisition of \mbox{Twitter/X} \\       
Platform usage (frequency)                  &       Comments on how often they use \mbox{Twitter/X} \\               
Platform usage (purpose)                    &       Comments on what they use \mbox{Twitter/X} for \\            
Toxicity definition                         &       Participant's definition of what constitutes toxicity (could be specifically in the context of a tweet reply) \\       
Relevance definition                        &       Participant's definition of what constitutes relevance specifically in the context of a tweet reply \\          
Engagement threshold (identity)             &       Participant's threshold or discussion of others' thresholds for seeing potentially harmful content based on identity \\                   
Engagement threshold (past experience)      &       Participant's threshold or discussion of others' thresholds for seeing potentially harmful content based on past or potential experiences \\                           
Engagement threshold (newsroom)             &       Participant's threshold or discussion of others' thresholds for seeing potentially harmful content based on newsroom standards \\
Engagement threshold (beat)                 &       Participant's threshold or discussion of others' thresholds for seeing potentially harmful content based on their beat \\               
Newsroom support                            &       Discussions of support to journalists provided by newsroom regarding abuse \\       
Other attack vectors                        &       Discussions of other ways that journalists have been or might be abused beyond Twitter replies \\           
Motivation of attackers                     &       Discussions of why an adversary has targeted them or executed an attack \\           
Personalizing classifiers                   &       Discussions of giving \SYSTEM feedback to improve its toxicity and relevance classifications \\
\SYSTEM extensions                          &       Discussions of ways to extend the tool to improve its utility (e.g. an additional feature, an additional application context) \\
Flagging content                            &       Content types that participant's want treated separately (i.e., motivated/persistent harasser, physical threat) \\
\bottomrule 
\caption{\textbf{Codebook for user testing} ---%
    Codebook for categorizing the different themes that emerged from our user testing study with eight journalists.}
\label{a:user_test_codebook}
\end{longtable}

\end{document}